\documentstyle[12pt,epsf,rotate,graphicx]{article}
\def\beq#1{\begin{equation}\label{#1}}
\def\eeq{\end{equation}}
\def\beqa#1{\begin{eqnarray}\label{#1}}
\def\eeqa{\end{eqnarray}}

\def\comment#1{\relax}

\def\spose#1{\hbox to 0pt{#1\hss}}
\def\simlt{\mathrel{\spose{\lower 3pt\hbox{$\mathchar"218$}}
     \raise 2.0pt\hbox{$\mathchar"13C$}}}
\def\simgt{\mathrel{\spose{\lower 3pt\hbox{$\mathchar"218$}}
     \raise 2.0pt\hbox{$\mathchar"13E$}}}
\def\simpropto{\mathrel{\spose{\lower 3pt\hbox{$\mathchar"218$}}
     \raise 2.0pt\hbox{$\propto$}}}

\makeatletter
\def\eqalign#1{\null\,\vcenter{\openup\jot\m@th
  \ialign{\strut\hfil$\displaystyle{##}$&$\displaystyle{{}##}$\hfil
      \crcr#1\crcr}}\,}
\def\eqalignleft#1{\null\,\vcenter{\openup\jot\m@th
  \ialign{\strut$\displaystyle{##}$\hfil&$\displaystyle{{}##}$\hfil
      \crcr#1\crcr}}\,}
\makeatother

\begin{document}
\pagestyle{empty}

\hfill{\sf Hot Points in Astrophysics}

\hfill{\sf JINR, Dubna, Russia, August 22-26, 2000}

\vspace*{2.cm}

\begin{center}

{\bf The 35-day cycle in Her X-1\\ as observational
appearance of freely precessing neutron star\\ and 
forcedly precessing accretion disk}

\vspace*{0.2cm}

N.A.Ketsaris$^{1}$, M.Kuster$^{2}$, K.A.~Postnov$^{1}$, 
M.E.~Prokhorov$^{1}$, N.I.Shakura$^{1}$, 
R.Staubert$^2$, and J.Wilms$^2$

{${}^1$ \it Sternberg Astronomical Institute, Moscow State University, 119899
Moscow, Russia}

{${}^2$ \it Institut f\"ur Astronomie und Astrophysik, Waldh\"auser Str. 64, 
D-72076 T\"ubingen, Germany}

\end{center}

\vspace*{0.2cm}

\begin{abstract}

A careful analysis of X-ray light curves and pulse profiles of Her X-1
obtained over more than 20 years strongly evidences for free precession of a
magnetized neutron star with rotational axis inclined to the orbital plane
as a central clock underlying the observed 35-day period. 
Strong asymmetric X-ray illumination of the optical star atmosphere 
leads to the formation of gaseous streams 
coming out of the orbital plane and forming 
a tilted accretion disk around the neutron 
star. Such a disk precesses due to tidal forces and dynamical
action of gaseous streams from the secondary companion.
The locking of these torques with neutron star precession
makes the net disk precession period to be very
close to that of the neutron star free precession. 

\end{abstract}

\section*{Introduction}

Her X-1 is an accretion-powered 1.24-s X-ray pulsar in a binary system with
1.7-d orbital period \cite{Tananbaum_ea72}. Since its discovery in 1972,
this source has puzzled the astronomers by its unusual complex behavior.

The X-ray light curve of Her X-1 is shaped with 
a 35-day  period, consisting of  a main-on X-ray
state with a mean duration of $\sim 7$ orbital periods surrounded by two
off-states (also called low-on states) each of $\sim 4$ orbital cycles, and of a
a short-on state of smaller intensity with a typical duration of
$\sim 5$ orbits, and is certainly due to periodic obscurations of the
X-ray source by the disk. In this system, we luckily observe it under the
angle of about 89 degrees, and it is this nearly edge-on position of the
line of sight that allows us to study a lot of tiny features in the X-ray
light curve. This first of all relates to the so-called X-ray dips, short
drops in the observed X-ray intensity accompanied by significant spectral
changes, which are observed each orbit after the turn-on. X-ray dips can be
explained by different models, but undoubtedly they are produced by
occultation of the central source by accreting material.
  
It is generally accepted that the 35-day period is due to counter-orbital
precession motion of a tilted accretion disk around the central neutron star
\cite{Gerend&Boynton76, Shakura_ea99}. Even earlier,
immediately after the discovery of Her X-1, it has been suggested that free
precession of the central neutron star causes the observed long-term
modulation with 35-day period 
\cite{Brecher72}. 
Some evidence for free precession indeed has been obtained from X-ray
pulse observed with EXOSAT satellite 
\cite{Truemper_ea86, Kahabka89}.

The observed stability of the 35-day period over many years would be
surprising for purely tidal precession of the accretion disk only, so the
need for an underlied clock mechanism was requested. This could be free
precession of neutron star, but at that time the locking between the neutron
star free precession period and tidal period of the accretion disk was
considered as a judicious assumption waiting for its physical grounds.

As was understood already shortly
after the beginning of studies of Her X-1, the accretion disk 
may be twisted. During the counter-orbital precession of such a disk 
the outer parts of the disk open the central X-ray source while the inner 
parts of the disk occult the X-ray source \cite{Boynton78}. 
Moreover, a hot 
rarefied accretion disk corona may exist around its central
parts.  This makes the ingress to and egress from main-on and short-on states
asymmetric. The opening of the X-ray source with a rapid increase
of X-ray intensity is accompanied  by a notable 
spectral changes which evidences for the presence of a strong absorption, 
whereas the decrease in X-ray intensity occurs more slowly
and without appreciable spectral changes
\cite{Giacconi et al. 1973}. 

One of the intriguing observational facts is that the X-ray source always
turns on near orbital phases $\phi_{orb}\simeq 0.2$ or $0.7$. Such a
behaviour has been explained by 
\cite{Levine and Jernigan (1982), Katz et al.
(1982)} by the accretion disk wobbling twice the orbital period due to tidal
torques. Indeed, it is at these orbital phases that the disk angle
inclination changes most rapidly.

Another notable feature is that the duration of 
successive 35-day cycles is as a rule 20, 20.5, or 21 orbital cycles
\cite{Staubert et al. 1983}. This behaviour has been confirmed by most 
recent RXTE observations \cite{Shakura_ea98}.

Even more enigmatic features observed are sudden decreases in X-ray flux
(X-ray dips) which are accompanied by significant spectral changes. They
have been observed by many X-ray satellites (see \cite{Shakura_ea99} for
references). X-ray dips are commonly separated into three groups:
pre-eclipse dips (P), which are observed in the first several orbits after
X-ray turn-on (up to 7 in main-on and up to 5 in short-on states) and march
from the eclipse toward earlier orbital phase in successive orbits;
anomalous dips (A), which are observed at $\phi_{orb}\sim 0.45-0.65$;
post-eclipse recoveries (R), which are occasionally observed as a short
delay (up to a few hours) of the egress from X-ray eclipse in the first
orbit after turn-on.

Changes in the Her X-1 pulse profile with the phase of a 35 day cycle have
been found in many observations (see \cite{Deeter_ea98, Scott_ea00} and references therein). As we will show below, to explain such profile
variation, emission region on the neutron star surface should have a complex
shape: in addition to the canonical magnetic poles, luminous rings around
the magnetic poles appear (as first discussed in \cite{Shakura_ea91,
Panchenko&Postnov94}. Due to free precession, these regions change
position with respect to the line of sight producing the observed slow pulse
shape evolution with 35-day cycle phase. Rapid pulse profile changes, which
are observed at the end of main-on stage, are due to occultation of the
neutron star surface by the precessing accretion disk. Soft X-ray sine-like
component of the pulse profile remains the only observed in low states, when
the neutron star surface is totally screened by the accretion disk. We will
show that this component is due to reprocessing of X-ray emission by the
innermost parts of the warped twisted accretion disk.
  
\section*{Evidence for free precession of neutron star
from X-ray pulse profiles}

As is well known, free precession of a non-spherical body changes the angle
between a given point on the body's surface and angular momentum vector if
none of the axes of inertia coincides with the angular momentum. As a
result, magnetic poles, which are the sources of X-ray emission, will
migrate with the precession phase, causing modulation of X-ray emission
observed by a remote observer.

In the case of free precession of a neutron star with strong magnetic field
surrounded by a diamagnetic thin accretion disk, the magnetospheric torques
applied to the disk will also change with the precession phase (see 
\cite{Lai99}
and references therein). The same torque but with the opposite sign will be
applied from the disk to the neutron star through the magnetic field. The
value and sign of this torque, averaged over the rapid neutron star
rotation, are both dependent on the obliquity angle $\theta$ between the
angular momentum and the magnetic pole. There is a critical value of this
angle, $\theta_c=54^o44''$, at which the average torque vanishes and the
innermost parts of the disk remain unperturbed. If $\theta>\theta_c$, the
inner disk acquires a twisted (helical) shape. The disk tends to keep in the
equatorial plane of the rotating neutron star. 
If $\theta<\theta_c$, the disk also tends to keep with the neutron star
equator, but with opposite helicity. 

The disk-magnetospheric interaction is a very complex and ill-understood, 
and in the case of neutron star free precession it seems likely that 
the rotational axis of the neutron star keeps oblique with respect 
to the orbital plane. The initial orbital inclination of the neutron star
axis can be caused by asymmetric supernova kick. However, accreted 
matter from the disk brings angular momentum which secularily tends to 
align the neutron star rotational axis with the orbital one. During 
free precession it is possible to incline the neutron star rotation axis 
by magnetospheric torques. Due to the helicity of the inner accretion 
disk (disk is not planar!), the neutron star spin axis tends to lie
into the orbital plane at $\theta<\theta_{cr}$ and vice versa.
The resulting inclination angle depends on the sum of differentials
$\Delta \theta$ caused by both magnetospheric torques and angular momentum
of accreted matter braught to the neutron star over precession cycle.         

\begin{figure}
\centerline{
\epsfysize=0.6\vsize
\rotate{\epsfbox{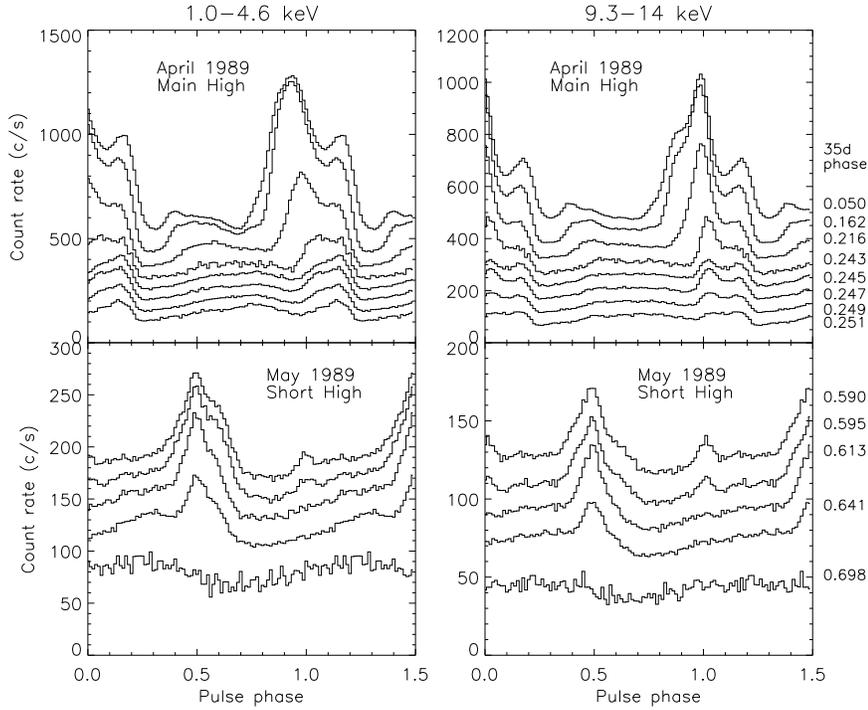}}
}
\caption{{\it Ginga} X-ray pulse profiles 
from paper \protect\cite{Scott_ea00}}
\label{pulses}
\end{figure}

In the simplest case of emission from magnetic poles only, the pulse profile
would consist of one or two pulses, depending on the visibility conditions
of the poles by the observer. However, in Her X-1 the pulse profile has a
more complicated shape (see Fig. \ref{pulses}). It consists of several peaks
that change first slowly and then rapidly with the precession phase. To
explain such an evolution, one needs to introduce a ring-like structures
around magnetic poles (for example, during a complex magnetic field
structure near the neutron star surface, \cite{Shakura_ea91}. Such
ring-like structures may be a natural result of a slow creeping away of
amorphous, well-conducting matter stored near the poles in the course of
accretion of plasma. During this process, the magnetic field diffuses
through the matter due to Ohmic dissipation of currents, so one may
simultaneously have several such structures of different sized around the
pole, and they well may have an elliptical, off-center shape around the
poles. The glowing of such rings is due to separation of the accreting flow
above the neutron star surface into different flows. In fact, these rings
can be describe as additional magnetic moments of the opposite sign relative
to the main magnetic moment of the neutron star \cite{Shakura_ea91}. As a
result, at a distance of several radii above the magnetic poles, the zone
appears with practically zero net magnetic field. It is in these zones where
the accretion flux separation occurs. Clearly, the most pronounced emission
will be from the rings most close to the poles, and the analysis of pulse
profiles of Her X-1 shows that there is two such rings around each pole.

In the main-on state of Her X-1 ($\phi_{pr}=0\div 0.24$) the magnetic poles
and the rings lie closer to the rotational axis, and in the short-on state
($\phi_{pr}=0.5\div 0.7$) they are closer to the rotational equator, and
moreover cross it during the precessional motion (cf. relative intensity
of the poles at phases 0.6-0.7 and in Fig. \ref{pulses}). 

Only a sine-like shape of the X-ray pulse remains visible  starting from
$\phi_{pr}\approx 0.26$ in the end of main-on state, and at
$\phi_{pr}\approx 0.66$ in the end of low-on state \cite{Deeter_ea98}.
Notably, these sine-like components are phase-shifted by 
180 degrees, which is
a natural consequence of precessional turn of the warped parts of inner
accretion disk.

\section*{ASM RXTE analysis of 35-day light curve}

\begin{figure}[t]
\centerline{
\epsfysize=0.5\vsize
\epsfbox{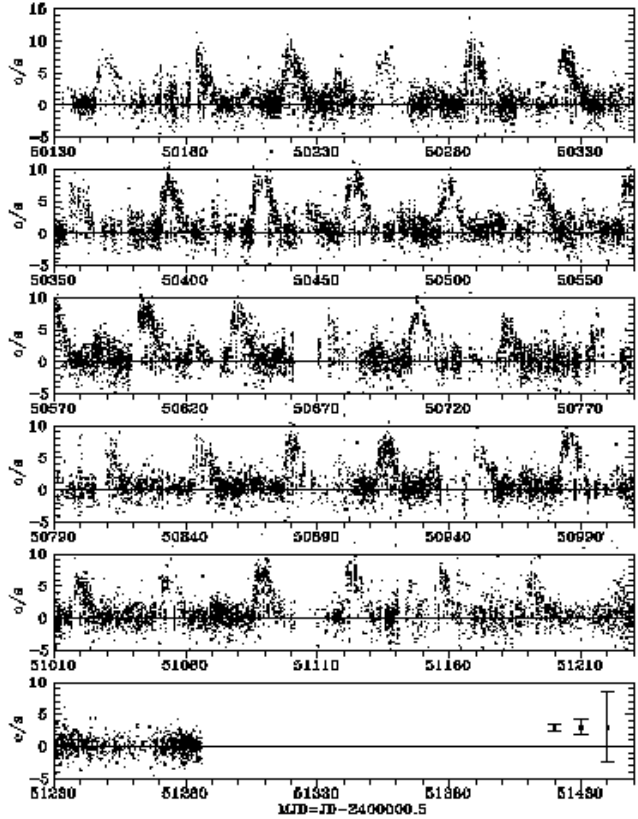}
}
\caption{ASM RXTE data}
\label{ASM}
\end{figure}

All-Sky Monitor onboard  RXTE satellite  \cite{Levine_ea96}
has obtained a lot of data on Her X-1.  
The data archive contains X-ray (2-12 keV) count rates averaged over
predominantly 90-s time intervals started from MJD 50087.
This monitoring revealed that the 35-day modulation stopped 
at the end of March of 1999, the source entered into an anomalous
low state which continues at present. Before this anomalous state, 
32 full 35-day cycles were continuously observed (Fig. \ref{ASM}). 
Below we present the updated analysis of these data (see 
\cite{Shakura_ea98} for early study).

Despite some gaps in the data, the archive is especially useful in
reconstructing the mean X-ray light curve by means of superposing many
cycles with account of turn-on times phases. In addition, it is possible to
determine the turn-on times of 35-day cycle with a good accuracy. This
accuracy is limited by the RXTE observations (the maximum UHURU count rate
from Her X-1 was about 100 counts per second while that of RXTE is about
$\le 10$ counts per second), nevertheless one may clearly distinguish around
which orbital phase, 0.25 or 0.75, each cycle was turned-on.

\begin{figure}[t]
\centerline{
\epsfxsize=0.6\hsize
\epsfbox{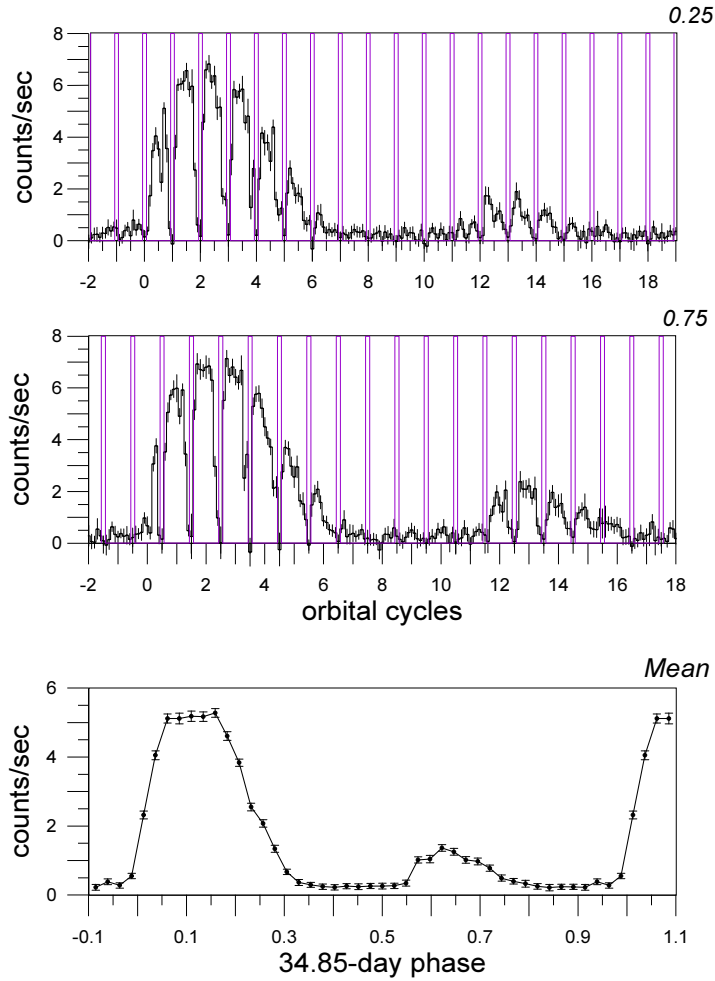}
}
\caption{Mean X-ray light curves of Her X-1: for 0.25 cycles (upper panel),
0.75 cycles (middle panel), and for all RXTE cycles (without eclipses).}
\label{lcurves}
\end{figure}

The analysis reveals that 19 cycles turned on near binary phase 0.25, 
and 13 near phase 0.75 (see \cite{Ketsaris_ea00} for more detail and
description of method used). The folded light curves for 0.25 and 0.75
cycles, as well as the mean 35-day light curve are presented in 
Fig. \ref{lcurves}.
Qualitatively, both main-on and short-on stages are seen. During the
main-on stage, the source reaches the peak count rate in 1.5 orbits,
stays at maximum for $\sim 2$ orbits, and progressively fades
to minimum during 3.5 orbits. The beginning of the short-on stage is
separated by 4.5 orbits from the end of the main-on. The duration of
the short-on state is about 4.5 orbits.

Note that the form of the short-on stage is similar to that of the main-on:
a rapid increase and slow decrease of count rate. The main-on state is
separated by the similar 4.5 orbits' off-state from the end of the
short-on state. The short-on stage decreases more slowly than
the main-on, so it is difficult to say where the second off-state
really starts. 
A weak X-ray glow
persistent during off-states is possibly due to scattering in the
accretion disk corona.

The anomalous dip is clearly seen for 0.25 cycles during the
first orbit after the main turn-on and is practically invisible at
0.75 cycles. For both types of cycles, pre-eclipsing dips demonstrate
identical behaviour -- they march from the eclipse toward earlier orbital
phase in successive orbits. No post-eclipse recovery was found in the
main-on state for 0.75 cycles.

\section*{Origin of X-ray dips}

\begin{figure}[t]
\epsfxsize=\textwidth
\epsfbox{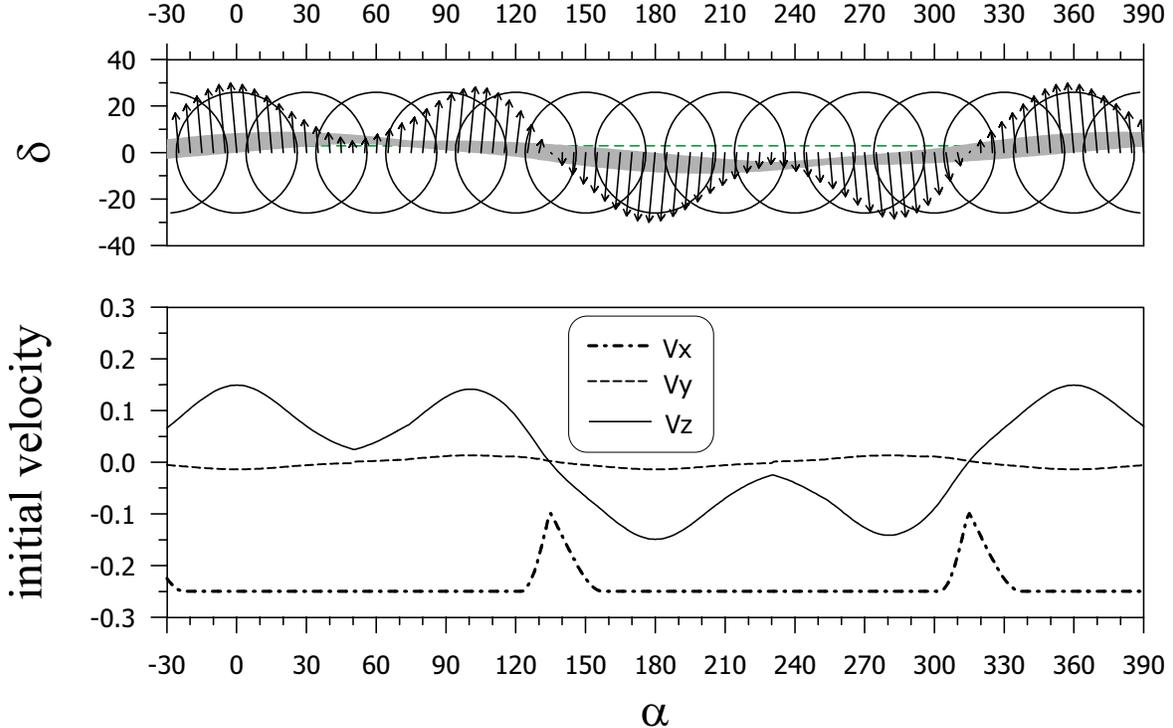}
\caption{Upper panel: the passage of HZ Her through the
shadow formed by a  tilted twisted 
accretion disk. The coordinates $\alpha$ and $\delta$
count from the line of nodes of the middle part of the 
disk along the orbital motion and from
the orbital plane, respectively. The dashed line indicates the
position of an observer inclined by $i=89^o$ to the
orbit. The arrows show the projection of the initial
accretion stream velocity on the plane
YZ perpendicular to the orbit. Bottom panel: The initial stream
velocity components at the $L_1$ point in units of the relative 
orbital velocity 270 km/s.}
\label{shadow}
\end{figure}

In our model, a tilted twisted counter-orbitally precessing accretion disk
eclipses the central X-ray source between the main-on and short-on states.
We model the disk by three parts: the outer disk, which is inclined 
by an angle of $\sim 5^o$ relative
to the orbital plane, the intermediate disk with approximately 
the same inclination as the outer disk but turned counter orbital 
rotation by an angle of
80 degrees, and the innermost disk whose orientation and
inclination are determined by magnetospheric torques. The disk 
ends at about the corotation radius $R_c=(GM/\omega^2)^{1/3}$, where
$\omega=2\pi/P$ is the neutron star angular rotation frequency and
$M$ its mass. The outer and intermediate disks counter precess as a solid
body, while the innermost disk changes orientation: in the main-on 
its inclination is about 9 degrees with a helicity angle of $\sim 9$ degrees
with respect to the intermediate disk, and in the short on the helicity 
is opposite with the same angle $-9$ degrees. Such variations are caused by 
changes in the angle $\theta$ between the neutron star angular momentum
and magnetic dipole axis. At $\theta=\theta_{cr}$ magnetospheric 
torques vanish and the innermost disk coincides with the intermediate one.
This takes place somewhere at low-on states (where the neutron
star surface is screened by the disk), when the magnetic pole 
goes up and down in due course of precession.  

Such a disk produces an appreciable shadow and the
optical star periodically enters this shadow in its orbital motion (Fig.
\ref{shadow}). 
The shadowed region is such that not all the optical star surface is
screened by the disk -- there always should exist areas illuminated by the
X-ray source with a photospheric temperature of 15,000-20,000 K whereas
photospheric temperature of the unheated regions is as low as $\sim 8,400$
K. Even higher temperatures (up to $10^6$ K) due to soft X-ray absorption by
heavy elements are attainable in the chromospheric layers over the
photosphere.

Such a high temperature induces matter outflow from the optical star which
would lie in the orbital plane in the absence of the shadow. 
When the shadow appears, a powerful pressure gradients emerge in the
chromospheric layers near the boundary separating illuminated and obscured
parts of the optical star, which initiates large-scale motions of matter
near the inner Lagrangian point $L_1$ with a large velocity component
perpendicular to the orbital plane \cite{Arons73, Katz73}. Such a shadow
will periodically modulate the matter outflow rate $\dot M$, which is
dramatically reduced when the $L_1$ point is deep inside the shadow, and
rises rapidly to a maximum at the moment when the shadow edge intersects the
$L_1$ point. Clearly, the picture repeats twice over the 
synodal orbital period.

\begin{figure} 
\epsfxsize=\columnwidth 
\epsfbox{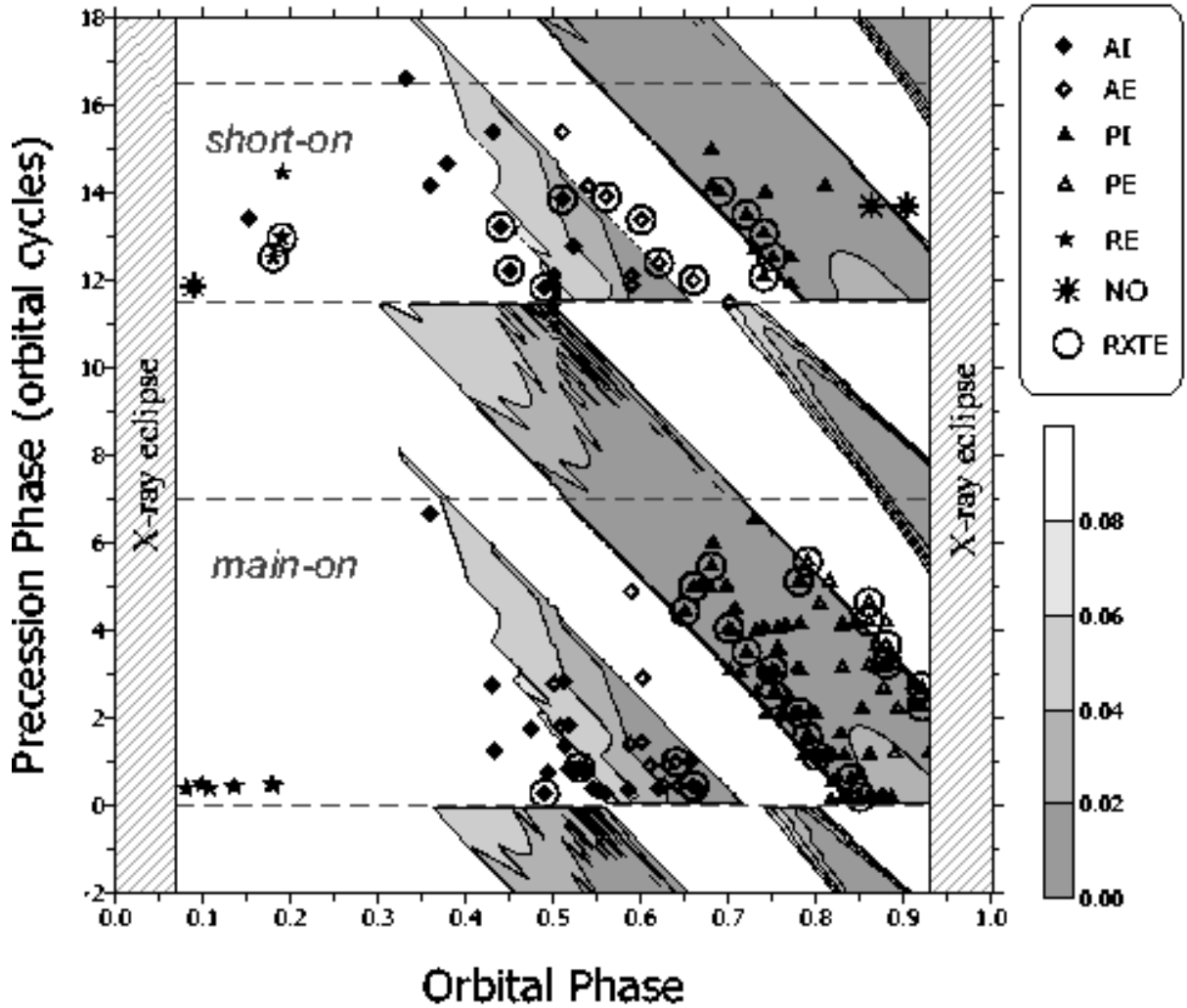} 
\caption{The part of
the plane $\phi_{orb}-\phi_{prec}$ with the observed
pre-eclipse and anomalous 
X-ray dips in the main-on and short-on states. Quadrangles and
triangles indicate ingress to (PI) and egress from (PE) the pre-eclipse dips.
RXTE data are encircled.
The calculated pre-eclipse dips and short anomalous dips arise when
the accretion stream intersects the line between the X-ray source and the
observer before entering the disk. Different contours correspond to
different minimal distances (0.02, 0.04, 0.06, 0.08) between the stream centre and
the line of sight.}
\label{dips}
\end{figure}

Thus, matter flows non-coplanar with the orbital plane emerge and supply 
the accretion disk with  angular momentum non-parallel to the orbital one.
Depending on the initial velocities, these streams may even increase the
disk tilt to the orbit. Such streams coming out of the $L_1$ point intersect
the line of sight of the observer at some orbital phases shortly before the
X-ray eclipse and shift slowly toward earlier phases as the precession
progresses. This is exactly the behaviour of the pre-eclipse X-ray dips
observed. The streams intersect the line of sight at other orbital phases
as well and thus give rise to the type I anomalous dips.

\begin{figure}[t]
\centerline{
\epsfxsize=0.8\hsize
\epsfbox{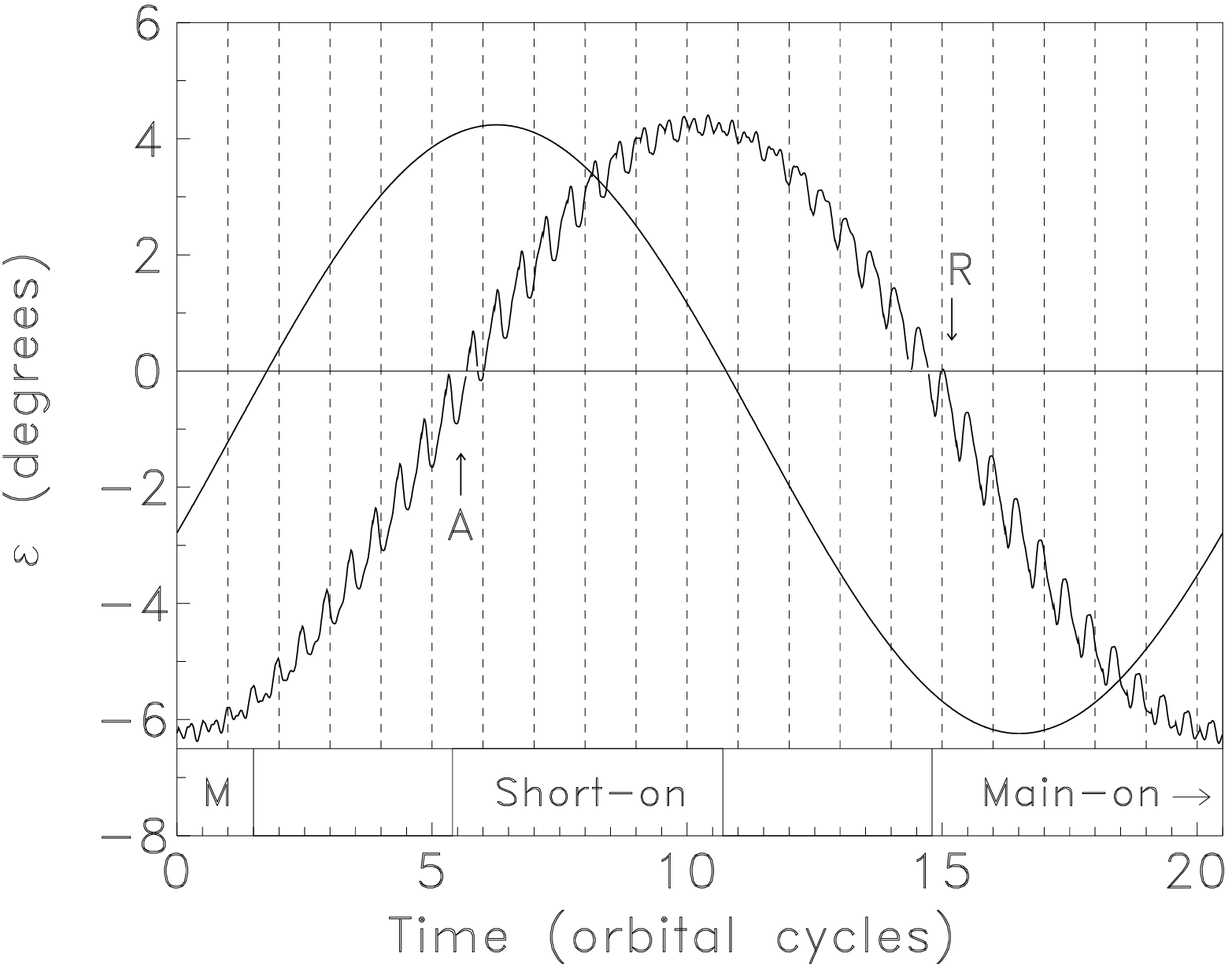}
}
\caption{Angle $\epsilon$ between 
the line of sight and the outer accretion disk plane
(the jagged line) and intermediate disk plane
(the smooth line) as a function of time (in 
orbital cycles). Small veridical arrows indicate
anomalous X-ray dips of type II (A) and
post-eclipse recovery (R).}  
\label{wobling}
\end{figure}

The problem of matter outflow from an asymmetrically illuminated stellar
atmosphere is essentially three-dimensional and requires sophisticated
numerical calculations. To obtain the pre-eclipse dips in a simplified
model, we calculated non-planar ballistic trajectories of particles ejected
from the point $L_1$.  Some trial functions for
the initial outflow velocity components are used 
(see \cite{Ketsaris_ea00} fro more detail).

Before colliding with the disk,  non-planar streams intersect the line
connecting the observer and central X-ray source thus absorbing some X-ray
flux. We identify these events with the pre-eclipse and type I anomalous
dips. In Fig. \ref{dips} 
we plot the calculated and observed dip positions on the
$\phi_{orb}-\phi_{pr}$ plane. The contours include the calculated
dip positions for different minimal distances between the stream centre and
the line of sight (0.02, 0.04, 0.06 and 0.08). 
The stream generated when the star
again enters the X-ray illuminated sector must also intersect
the line of sight during the short-on state
$\phi_{pr}=0.35-0.65$. This gives rise to the pre-eclipse dips during the
short-on state as well, as indeed observed 
\cite{Jones&Forman76,  Ricketts_ea82,  Shakura_ea98}.
 
Fig. \ref{dips} 
demonstrates qualitative agreement between the observed and
calculated dip positions.
 Note that in our model additional
pre-eclipse X-ray dips can appear at the end of 
the main-on and short-on states. By varying slightly 
the parameters it is easy to obtain the merging 
of the end of the pre-eclipse dip with the beginning of
such "third" dips, so that the central X-ray source
remains unobservable until the beginning of orbital X-ray eclipse.
Indeed, {\it Ariel-V} data \cite{Ricketts_ea82} clearly demonstrated
the absence of egress from pre-eclipse dips at several short-on states!

\begin{figure}[t]
\centerline{
\epsfxsize=0.7\hsize
\epsfbox{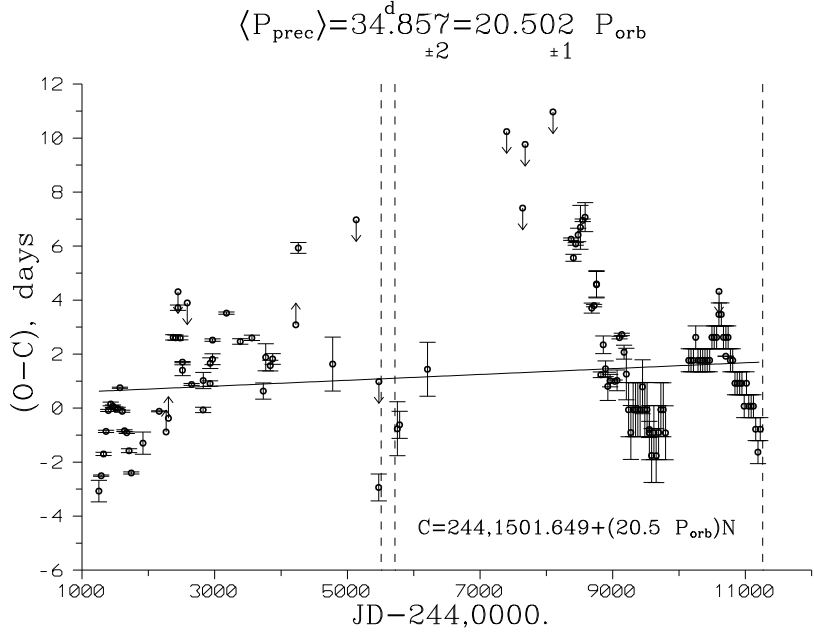}
}
\caption{Observed minus Calculated (O-C) diagram for 
turn-ons of Her X-1 over 28 years. The dashed curves indicate 
9-months' anomalous low state of Her X-1 in 1983-84, and the beginning
of the current anomalous low state (March 1999).}
\label{O-C}
\end{figure}

Unlike the pre-eclipse dips and type I anomalous dips, the anomalous dips of
type II and post-eclipse recoveries are formed by another mechanism. The
vector of the outer disk angular momentum moves along a precession cone and
undergoes an oscillating (wobbling) motion twice the synodical orbital
period. The wobbling arises due to joint action of streams and
tides (see  Fig. \ref{wobling} and 
 \cite{Shakura_ea99} for more detail). The wobbling amplitude
is higher in short-on than in main-on state, which can be a possible
reason for more strong appearance of post-eclipse recoveries 
observed in short-on.

\section*{Discussion and conclusion}

The model of the unique binary system Her X-1 invokes both a free precessing
neutron star and a tidally precessing accretion disk. Clearly, in general
case the neutron star free precession period and accretion disk precession
period should be different. In Her X-1, X-ray source opens by the precessing
outer parts of the disk. As we noted above, the turn-ons always occur
randomly in 20, 20.5, or 21 orbits. The result of this random process is
that time delay in some turn-ons relative to the mean ephemeris can be as
long as 3-3.5 orbits (see Fig. \ref{O-C}). Nevertheless, the mean period is
$\langle P_{prec}\rangle=34^d.875\pm 0^d.002=20.502\pm 1\times P_{orb}$, 
which we believe to be the neutron star free precession period. 
The closeness of this period to 20.5 orbits is quite spurious
(cf. the closeness of apparent angular diameters of Sun and Moon).
Actually, this period is not the strict one, since the neutron star
can undergo quakes etc. We believe that the locking 
of the disk precession
with the neutron star free precession is due to dynamical action of
the gaseous stream forming the disk: when it tends to go ahead, 
the streams brake its precessional motion, and vice versa.

\vskip\baselineskip

{\it Acknowledgments}
The work was supported by the Grant "Universities of
Russia", No5559, Russian Fund for Basic Research through Grant No
00-02-17164, and NATO grant PST.CLG 975254. 

\vskip\baselineskip

\end{document}